# Contribution of the Hall effect to radial electric field and spontaneous/intrinsic rotation in tokamak core plasmas

A.B. Kukushkin[1,2], M.G. Levashova[1]

[1] *National Research Center "Kurchatov Institute", Moscow, Russia*
[2] *National Research Nuclear University MEPhI, Moscow, Russia*

The Hall effect, defined as the separation of electric charges of opposite sign when they move in a magnetic field, is suggested to contribute substantially to the observed negative radial electric field $E_r$ in the core plasma in tokamaks and, respectively, to the spontaneous/intrinsic rotation of plasma. A simple way to evaluate the Hall effect contribution to the $E_r$ value, using the independently measured space distributions of magnetic field and plasma rotation velocity, is suggested. The estimates of the effect for experimental data from the TM-4 and T-10 tokamaks suggest that the above phenomena in tokamaks should be described in the framework of the two-fluid magnetohydrodynamics.

## 1. Introduction.

A detailed study of the role of the Hall effect in plasmas began with the study in [1] — within the framework of two-fluid magnetohydrodynamics (MHD) — of the effect of the frozenness of magnetic field mainly in the electronic component of plasma. This effect is important for stationary plasma flows (see the review [2]), and it plays a dominant role in the fast processes in plasmas (plasma open switches, magneto-inertial confinement of plasma in the pulsed discharges, including the Z-pinches and plasma foci) and widely studied in the literature (see, e.g., the review [3]). For example, the use of the model of magnetic field dynamics in plasmas, which simultaneously takes into account the transfer of the (almost frozen into electrons) magnetic field with the current velocity of electrons and the diffusion of magnetic field in plasma, allowed to build qualitative models of the enhanced-rate (in comparison with the usual diffusion) propagation of magnetic field in plasmas due to the Hall effect. This approach allowed [4] to describe the observed phenomena of the enhanced-rate propagation of the electric current shell along the anode in the plasma focus discharges of the Filippov type. Prior to this, analytical model [5] for the dynamics of the magnetic field and plasma near the anode surface and the two-dimensional numerical simulations [6] of plasma dynamics in the entire volume of the discharge chamber existed for this phenomenon.

In recent years, great progress has been made in the direct measurements of the radial electric field, $E_r$, using the heavy ion beam probe (HIBP) diagnostics (see, e.g., [7]). This allows one to check the theoretical predictions derived by various methods. For instance, in [8] the behavior of $E_r$ in the tokamak core plasma was qualitatively explained by neoclassical theory, and in the peripheral plasma – in the two-fluid model of turbulent transport. The origin of the negative field $E_r$ in the core plasma in the neoclassical model is explained by the escape of ions from the core plasma, while the predictive modeling of the force equilibrium was carried out in [8] using the ASTRA code simulations in the framework of the single-fluid MHD.

The lack of the proper attention to the equation of force equilibrium for electrons (in other words, underestimation of the role of this equation and of the Hall effect described by it) is, in our opinion, a common feature of the works on the theory of $E_r$ and spontaneous plasma rotation in tokamaks. The focus has been made on the studies of various other effects in the framework of the single-fluid MHD models for plasma equilibrium (see for example, a brief overview of the status of plasma rotation studies in [9]). The description of $E_r$ in tokamaks is dominated by the approach where the plasma equilibrium is described by the single-fluid MHD, while the equation of equilibrium for the ions is used to determine the radial electric field from the measured values of ion pressure and toroidal and poloidal rotation velocities (see, e.g., [10]).

A simplified two-fluid MHD model, the "Hall MHD", was suggested in [11] and developed in [12] for the case of plasma rotation velocities much higher than the electric current velocity (in this case, $E_r$ is neglected). We consider here the opposite limit for the ratio of the above-mentioned velocities.

The progress in modelling the effects of finite macroscopic velocities on axisymmetric ideal equilibria is made with the two-fluid (ions and electrons) model of the equilibrium solver, the code FLOW2 [13], where the original approaches [2, 14] are implemented and extended. The problem of the origin and the value of radial electric field is not considered in [13].

Here, we show that the Hall effect may substantially contribute to the observed negative radial electric field $E_r$ in the core plasma in tokamaks and, respectively, to the spontaneous/intrinsic rotation of plasma. The principles of the model are outlined in Sec. 2, where a simple way to evaluate the Hall effect contribution to the $E_r$ value, using the independently measured space distributions of magnetic field and plasma rotation velocity, is suggested. The estimates of the effect for experimental data from the TM-4 and T-10 tokamaks are presented in Sec. 3. The conclusions are made in Sec. 4, where the necessity of the two-fluid MHD for describing the above phenomena is claimed.

## 2. A simple model for evaluating the contribution of the Hall effect to radial electric field in tokamaks

The equations for the electron and ion radial equilibrium in the two-fluid model have the well-known form, and — for negligible viscosity, deviation from electric charge neutrality, inertia and centrifugal force — may be written in the following form:

$$\frac{\partial p_e}{\partial r} = -en_e E_r - \frac{1}{c}j_{tor}B_{pol} + \frac{1}{c}j_{pol}B_{tor} + \frac{en_e}{c}(V_{tor}B_{pol} - V_{pol}B_{tor}), \tag{1}$$

$$\frac{\partial p_i}{\partial r} = en_e E_r - \frac{en_e}{c}(V_{tor}B_{pol} - V_{pol}B_{tor}), \tag{2}$$

where $p_e$ and $p_i$ are the electron and ion pressure, respectively; $n_e$ is electron density; $e$ is absolute value of electron charge; $j_{tor}$ and $j_{pol}$ are the toroidal and poloidal electric current density, respectively; $B_{tor}$ and $B_{pol}$ are the toroidal and poloidal magnetic field, respectively; $V_{tor}$ and $V_{pol}$ are the toroidal and poloidal hydrodynamic velocity of plasma (i.e. ions), respectively.

Here we consider the case when there is no substantial contribution to the plasma rotation from the external, auxiliary (with respect to the Ohmic mode of tokamak operation) sources like the neutral beam injection (NBI). Thus, we treat only the case where the plasma rotation is called spontaneous or intrinsic one. Under these conditions, the electric current velocity is much higher than the plasma rotation velocity, and the Ampere's force, i.e. the second and the third terms in Eq. (1), dominates over the contribution of the plasma rotation, i.e. the forth and the fifth terms in Eq. (1). This case is surely applicable to the Ohmic mode of tokamak operation.

In the literature, the value of electric field, which equates the Ampere's force, is called the Hall electric field,

$$\boldsymbol{E}_{Hall} = \frac{1}{en_e c}[\boldsymbol{j}, \boldsymbol{B}]. \tag{3}$$

For a quasi-steady-state radial electric field in tokamaks, it would be natural to redefine the Hall electric field,

$$\boldsymbol{E}_{Hall}^* = \frac{1}{en_e}\left\{\frac{1}{c}[\boldsymbol{j}, \boldsymbol{B}] - \frac{\partial p_e}{\partial r}\right\}, \tag{4}$$

because the pressure gradient weakens (saturates) the separation of the electric charges by the Lorentz force to give the net electric field.

In the single-fluid ideal plasma, the MHD velocity of plasma across magnetic field also produces the electric field due to the Lorentz force (equivalently, the plasma moves with the E-cross-B drift velocity in the crossed electric and magnetic fields). Therefore, the contribution of the plasma motion to Eq. (1) may be qualified as the electric field of the "ideal MHD" origin:

$$\boldsymbol{E}_{idealMHD} = -\frac{1}{c}[\boldsymbol{V}, \boldsymbol{B}], \tag{5}$$

where $\boldsymbol{V}$ is the plasma (i.e. ions) MHD velocity.

It follows from Eq. (1) that the poloidal magnetic field is responsible for the following effects:
- non-steady-state $B_{pol}$ produces the toroidal electric current (in the Ohmic mode, the poloidal magnetic field is pumped into the main chamber from the poloidal current coils);
- $B_{pol}$ compresses electrons (the pinch effect) and separates the electric charges because ions are magnetized much less than electrons (this separation produces radial electric field);
- $B_{pol}$ compresses the plasma with the (almost frozen-in) toroidal magnetic field and creates the poloidal electric current.

Note that all these effects are well recognized in the physics of magneto-inertially confined plasmas (cf. [3-6]).

The balance of the forces on electrons (for small ion MHD velocity as compared to that of electrons) is as follows:
- compression by the poloidal magnetic field (negative values of the force),
- repulsion by the compressed toroidal magnetic field (positive values of the force),
- repulsion by the pressure of the particles (diamagnetic effect, positive values of the force),
- repulsion due to an excess of negative electric charge (positive values of the force for negative values of electric field).

It follows from Eq. (2) that the above mechanism of generating the radial electric field should generate "spontaneous" rotation of ions in the crossed fields: radial electric and poloidal and toroidal magnetic ones (taking into account the ion pressure, i.e. ion diamagnetic drift). If the contribution of the "ideal MHD" electric field to Eq. (1) is relatively small (that is the case for plasmas without strong external sources of plasma rotation), Eq. (2) may serve for evaluating the plasma rotation velocity generated by the (directly measured or determined from Eq. (1)) radial electric field, rather than for evaluating the radial electric field of the *unknown* origin from the measured values of plasma rotation velocities and ion pressure.

### 3. Estimates for TM-4 and T-10 tokamaks

The model of Sec. 2 may be applied to evaluating the $E_r$ value from the measured values of electric current density, electron pressure and plasma rotation velocity. Here we try the model for the data [15] from the TM-4 tokamak and the data [8, 16] from the T-10 tokamak.

For estimates, we use the *canonical* profiles for the pressure and toroidal electric current (see [17] and references therein), taking $\alpha_T = 3$:

$$p_e(\rho) = \frac{p_e(0)}{(1 + \alpha_T \rho^2)^2} \quad (6)$$

$$j_{tor}(\rho) = \frac{j_{tor}(0)}{(1 + \alpha_T \rho^2)^2} \quad (7)$$

When evaluating the contribution of the total value of the Ampere's force (i.e. with account of the poloidal electric current), we assume that the profile of the magnetic-surface-average value of the total magnetic field is nearly flat (cf., e.g., Fig. 1a in [18] for the ITER scenario), so that this contribution approximately reduces to the value of $B_{pol}^2/(4\pi\rho)$.

For the TM-4 tokamak, we use the data [15]: minor radius a = 0.085 m, $T_e(0)$ = 0.5 keV, $n_e(0)$ = 3.0 $10^{19}$ m$^{-3}$, plasma toroidal electric current $I_p(a)$ = 0.025 MA, $B_{tor}(0)$ = 1.45 T, $V_{pol}(\rho=0.5)$ ~ 2 km/s (cf. Fig. 4 in [15]). The results of evaluating the terms in Eq. (1) are shown in Fig. 1, and the radial electric field, in Fig. 2.

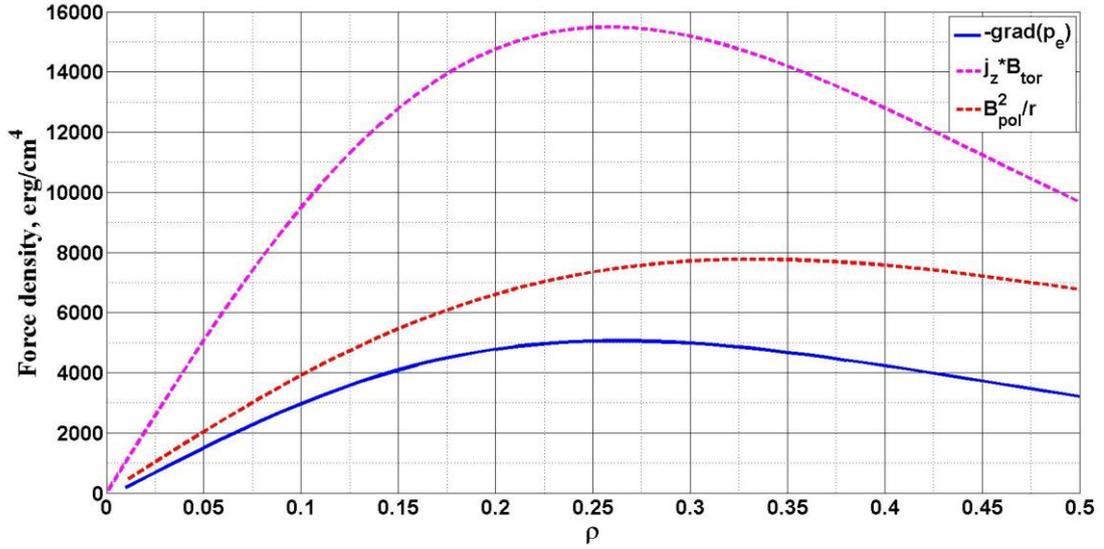

Figure 1. Comparison of the terms in the electron equilibrium equation, Eq. (1), as the functions of the normalized minor radius, for the TM-4 tokamak data [15]: toroidal electric current contribution to the Ampere's force (dashed magenta curve), the total Ampere's force (dashed red curve), electron pressure gradient (blue).

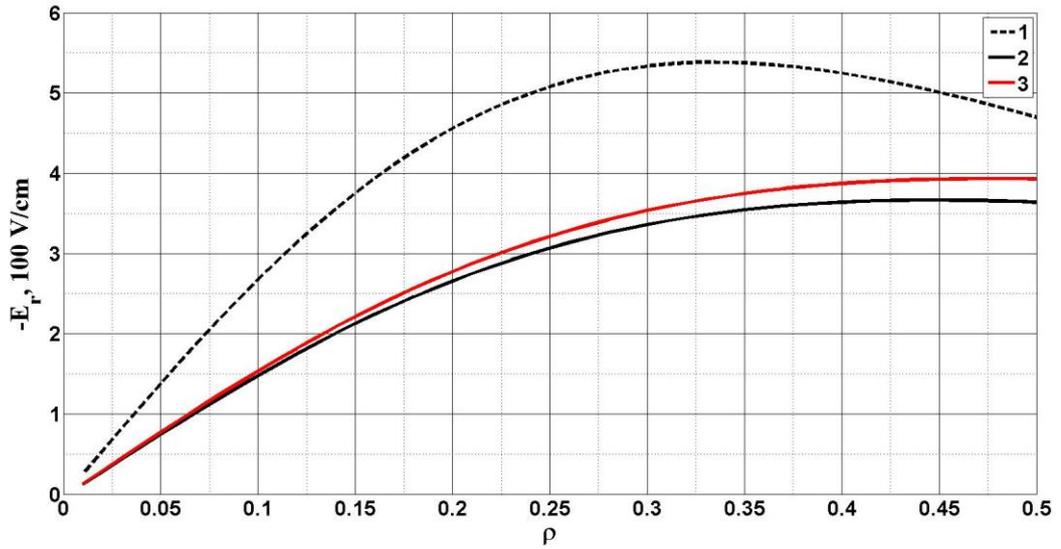

Figure 2. The profile of the radial electric field evaluated from Eq. (1) for the TM-4 tokamak data [15]: the effective Hall electric field (4) with account of toroidal electric current only (dashed black curve), the same for the total Ampere's force (black curve), the same with account of contribution (5), i.e. of all the terms in Eq. (1) (red).

It is seen that the roughly estimated value of radial electric field in the core plasma exceeds the values directly measured in [15] (~100 V/cm), being of the same order of magnitude.

For the T-10 tokamak, we use the data [8, 16]: minor radius a = 0.3 m, $T_e(0)$ = 0.8 keV, $n_e(0)$ = 4.0 $10^{19}$ m$^{-3}$, plasma toroidal electric current $I_p(a)$ = 0.15 MA, $B_{tor}(0)$ = 2.1 T, $V_{pol}(\rho=0.5)$ ~ 3 km/s (cf. Figs. 2 and 4 in [8]). The results of evaluating the terms in Eq. (1) are shown in Fig. 3, and the radial electric field, in Fig. 4.

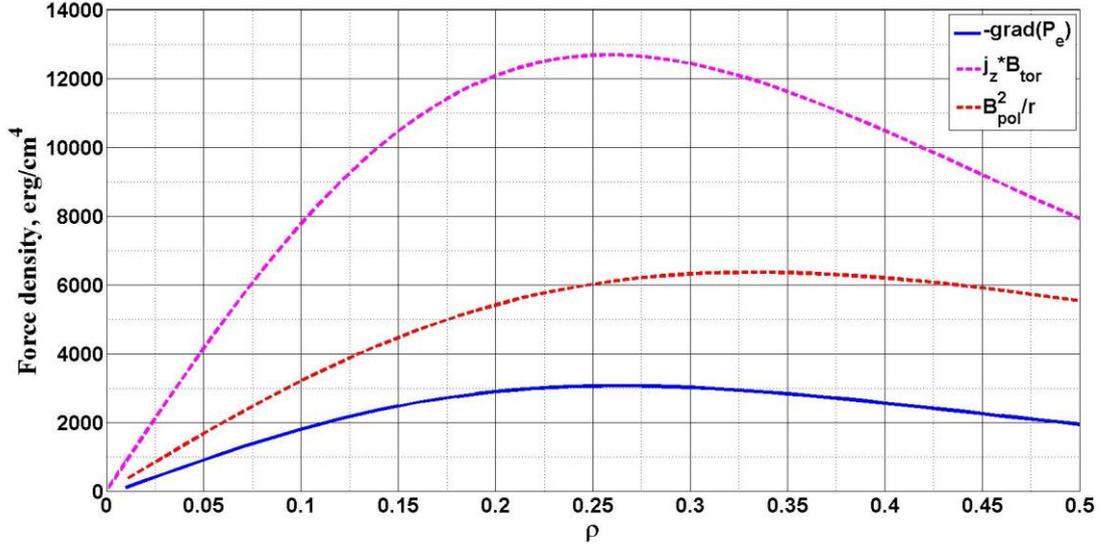

Figure 3. Comparison of the terms in the electron equilibrium equation, Eq. (1), as the functions of the normalized minor radius, for the T-10 tokamak data [8]: toroidal electric current contribution to the Ampere's force (dashed magenta curve), the total Ampere's force (dashed red curve), electron pressure gradient (blue).

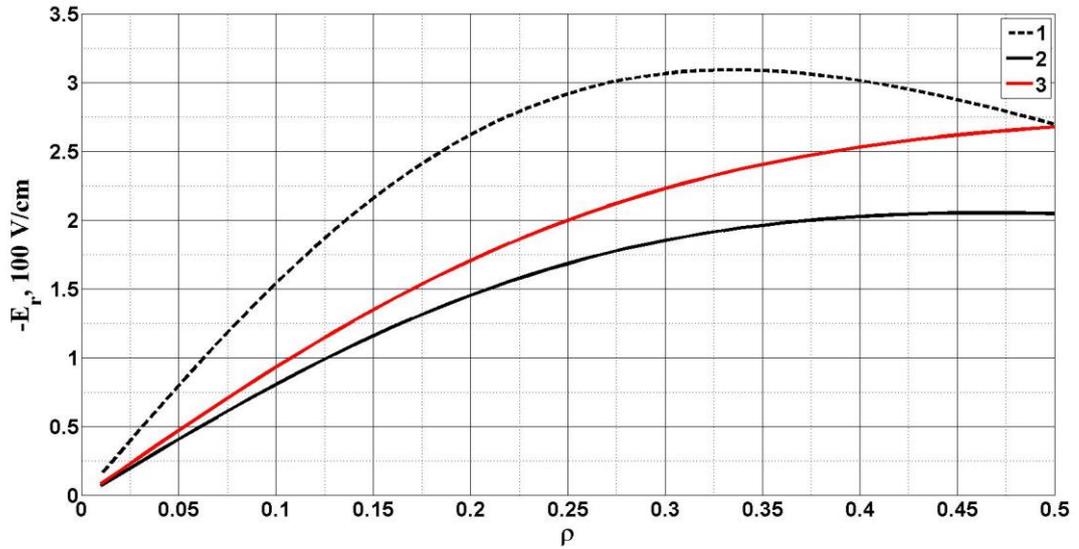

Figure 4. The profile of the radial electric field evaluated from Eq. (1) for the T-10 tokamak data [8, 16]: the effective Hall electric field (4) with account of toroidal electric current only (dashed black curve), the same for the total Ampere's force (black curve), the same with account of contribution (5), i.e. of all the terms in Eq. (1) (red).

It is seen that the roughly estimated value of radial electric field in the core plasma exceeds the values directly measured in [8, 16] (~100 V/cm), being of the same order of magnitude that the directly measured values and those simulated in [8] with the neoclassical model for the core plasma.

**4. Conclusions**

The present paper is aimed at
- formulating the principle of evaluating the contribution of the Hall effect to radial electric field $E_r$ and spontaneous/intrinsic rotation in tokamak core plasmas, in the case of negligible contribution of external sources of plasma rotation, like NBI,
- making the estimates of the Hall effect using the data from the TM-4 and T-10 tokamaks to illustrate possible status of the Hall effect in tokamaks.

We may draw the following conclusions.
1. The Hall effect, defined as the separation of electric charges of opposite sign when they move in a magnetic field, may contribute substantially to the observed negative radial electric field $E_r$ in the core plasma in tokamaks and, respectively, to the spontaneous/intrinsic rotation of plasma.
2. The estimates of the effect for experimental data from the TM-4 and T-10 tokamaks suggest that the above phenomena in tokamaks should be described in the framework of the two-fluid magnetohydrodynamics (MHD).

It follows, in particular, from the second conclusion that the existing approaches to explaining/predicting the values of radial electric field and plasma rotation velocity should be appended with using the **two-fluid** MHD model for plasma equilibrium.

**Acknowledgements.**

The present paper is dedicated to the 90-th anniversary of the outstanding plasma physicist Professor Alexey I. Morozov (1928-2009).

The authors are grateful to A.V. Melnikov for stimulating discussions, Yu.N. Dnestrovskij, L.G. Eliseev, V.A. Krupin, V.S. Mukhovatov, G.E. Notkin, K.A. Razumova, V.S. Strelkov, V.A. Vershkov, for helpful discussion of the present work, all the participants of the Plasma Theory Division seminar and the Tokamak Experiment Division seminar at the NRC "Kurchatov Institute", for helpful discussions.